\documentclass{rspublic}

\begin{document}
\title[Cosmology with Varying Constants]{Cosmology with Varying Constants}
\author[C.J.A.P. Martins]{Carlos J. A. P. Martins}

\affiliation{
CAUP, R. das Estrelas s/n, 4150-762 Porto, Portugal,\\
Department of Applied Mathematics and Theoretical Physics, CMS,\\
University of Cambridge, Wilberforce Road, Cambridge CB3 0WA, UK, and\\
Institut d'Astrophysique de Paris, 98 bis Boulevard Arago,
75014 Paris, France}
\label{firstpage}
\maketitle

\begin{abstract}{Cosmology; Extra dimensions; Fundamental constants; Laboratory and astrophysical tests}
The idea of possible time or space variations of
the `fundamental' constants of nature, although not new,
is only now beginning to be actively considered by large numbers
of researchers in the particle physics, cosmology and astrophysics
communities. This revival is mostly due to the claims of
possible detection of such variations, in various different contexts and
by several groups. Here, I present the current theoretical motivations and
expectations for such variations, review the current observational status,
and discuss the impact of a possible confirmation of these results in our
views of cosmology and physics as a whole.
\end{abstract}

\section{Introduction}

One of the most valued guiding principles (or should one say beliefs?)
in science is that there ought to be a single, immutable set of laws
governing the universe, and that these laws should remain the same
everywhere and at all times. In fact, this is often generalised into
a belief of immutability of the universe itself---a much stronger
statement which doesn't follow from the former.
A striking common feature of almost all cosmological models throughout
history, from ancient Babylonian models, through the model of Ptolemy and
Aristotle, to the much more recent `steady-state model', is their immutable
character. Even today, a non-negligible minority of cosmologists still speaks
in a dangerously mystic tone of the allegedly superior virtues
of `eternal' or `cyclic' models of the universe.

It was Einstein (who originally introduced the cosmological constant as a
`quick-fix' to preserve a static universe) who
taught us that space and time are not an immutable arena
in which the cosmic drama is acted out, but are in fact part of the
cast---part of the physical universe. As physical entities, the
properties of time and
space can change as a result of gravitational processes.
Interestingly enough, it was soon after the appearance of General Relativity,
the Friedman models, and Hubble's discovery of the expansion of the
universe---which shattered the notion of immutability of the
universe---that time-varying fundamental constants
first appeared in the context of a complete cosmological model (Dirac, 1937),
though others before him (starting with Kelvin and Tait)
had already entertained this possibility.

From here onwards, the topic remained somewhat
marginal, but never disappeared completely, and even the Royal Society
organised a discussion on this theme about twenty years ago.
The proceedings of this discussion (McCrea \& Rees 1983) still make very
interesting reading today---even if, in the case
of some of the articles, only as a reminder
that concepts and assumptions that are at one point uncontroversial
and taken for granted by everybody in a given field can soon afterwards
be shown to be wrong, irrelevant or simply `dead-ends' that are abandoned
in favour of an altogether different approach.

Despite the best efforts of a few outstanding theorists, it took as usual
some observational hints for possible variations of the fundamental
constants (Webb \textit{et al.} 1999)
to make the alarm bells sound in the community as a whole,
and start convincing some previously sworn skeptics. In the past
two years there has been an unprecedented explosion of interest in this area,
as large as (or perhaps even larger than) the one caused by the evidence
for an accelerating universe provided by Type Ia supernovae data. On one
hand, observers and experimentalists have tried to reproduce these results
and update and improve other existing constraints. On the other hand, a swarm
of theorists has flooded scientific journals with a whole range
of possible explanations.

Here I will provide a brief summary of the current status of this topic.
Rather than go through the whole zoo of possible models (which would
require a considerably larger space, even if I were to try to separate
the wheat from the chaff), I'll concentrate in the model-independent
aspects of the problem, as well as on the present observational status.
Towards the end, I'll provide some reflections on the impact of a future
confirmation of these time variations in our
views of cosmology and physics as a whole.

\section{On the role of the constants of nature}

The so-called fundamental constants of nature are widely regarded as
some kind of distillation or `executive summary' of physics.
Their dimensions are intimately related to the form and structure of
physical laws. Almost all physicists (and even engineers) will
have had the experience of momentarily forgetting the exact expression
of a certain physical law, but quickly being able to re-derive it
simply by resorting to dimensional analysis.
Despite their perceived fundamental nature, there is
no theory of constants as such. How do they originate?
How do they relate to one another? How many are necessary to
describe physics? None of these questions has a rigorous answer at present.
Indeed, it is remarkable to find that different people can have so widely
different views on such a basic and seemingly uncontroversial issue.
Duff \textit{et al.} (2002) has a very interesting discussion of this issue.

One common view of constants is as asymptotic states. For example, the
speed of light $c$ is (in special relativity) the maximum velocity
of a massive particle moving in flat spacetime. The gravitational
constant $G$ defines the limiting potential for a mass that doesn't form
a black hole in curved spacetime. The reduced Planck constant
$\hbar\equiv h/2\pi$ is the universal quantum of action (and hence defines a
minimum uncertainty). Similarly in string theory there
is also a fundamental unit of length (the characteristic size of the strings).
So for any physical theory we know of, there should be one such constant.
This view is acceptable in practice, but unsatisfactory in principle,
mainly because it doesn't address the question of the constant's origin.

Another view is that they are simply necessary (or should one say
convenient?) inventions, that is, they are not really fundamental
but simply ways of relating quantities
of different dimensional types. In other words, they are simply conversion
constants which make the equations of physics dimensionally homogeneous.
This view, first clearly formulated
by Eddington (1939) is perhaps at the origin
of the tradition of absorbing constants (or `setting them to unity', as
it is often colloquially put) in the equations of physics.
This is particularly common in areas such as relativity,
where the algebra can be so heavy that cutting down the number of symbols
is a most welcome measure. However, it should be remembered that this procedure
can not be carried arbitrarily far.
For example, we can consistently set $G=h=c=1$, but we can not set
$e=\hbar=c=1$ since then the fine-structure constant would have the value
$\alpha\equiv e^2/(\hbar c)=1$ whereas in the real world $\alpha\sim1/137$.

In any case, one should also keep in mind that the possible choices of
particular units are infinite and always arbitrary. For example,
the metre was originally defined as the distance between two scratch
marks on a bar of metal kept in Paris. Now it is defined in terms of a number
of wavelengths of a certain line of the spectrum of a ${}^{83}Kr$ lamp.
This may sound quite more `high-tech' and rigorous, but it doesn't really
make it any more meaningful.

Perhaps the key point is the one recently made by Veneziano in
(Duff \textit{et al.} 2002):
there are units which are arbitrary and units
which are fundamental at least in the sense that, when a quantity becomes
of order unity in the latter units, dramatic new phenomena occur.
For example, if there was no fundamental length, the properties of physical
systems would be invariant under an overall rescaling of their size,
so atoms would not have a characteristic size, and we wouldn't even be able
to agree on which unit to use as a `metre'. With a fundamental quantum unit of
length, we can meaningfully talk about short or large distances. Naturally we
will do this by comparison to this fundamental length.
In other words, `fundamental' constants are fundamental only
to the extent that they provide us with a way of
transforming any physical quantity (in whatever units we have chosen to
measure it) into a pure number whose meaning is immediately clear and
unambiguous.

Still, how many really `fundamental' constants are there? Note that some
so-called fundamental units are clearly redundant: a
good example is temperature, which is simply the average energy of a system.
In our everyday experience, it turns out that we
need three and only three: a length, a time, and an energy. However,
it is possible that in higher-dimensional theories (such as string theory,
see \S4) only two of these may be sufficient. And maybe, if and when the
`theory of everything' is discovered, we will find that even less than two are
required---again, refer to Duff \textit{et al.} (2002)
for a more detailed discussion.

\section{Standard cosmology: what we know and what we don't}

Cosmology studies the origin and evolution of the universe, and in particular
of its large-scale structures, on the basis of physical laws. By large-scale
structures I mean scales of galaxies and beyond. This is the scale where
interesting dynamics is happening today (anything happening on scales
below this one is largely irrelevant for cosmological dynamics).
The standard cosmological model, which was gradually put together
during the twentieth century, is called the `Hot Big Bang'
model. Starting from some very simple assumptions, it leads to a number of
predictions which have been observationally confirmed.

Three of these predictions are particularly noteworthy.
Firstly, there is Hubble's law---the fact that the
universe is expanding, and galaxies are moving away from each other with a
speed that is approximately proportional to the distance separating them.
Secondly, Big Bang Nucleosynthesis (BBN) predicts the relative primordial
abundances of the light chemical elements (which were synthesised in
the first three minutes of the universe's existence): roughly
$75\%$ Hydrogen, $24\%$ Helium and only $1\%$ other
elements. Last but not least, there is the Cosmic Microwave Background (CMB).
This is a relic of the very hot and dense early universe. By measuring photons
from this background coming from all directions, one finds an almost perfect
black body distribution with a present temperature
of only $2.725$ degrees Kelvin, corresponding to a present radiation density of
about 412 photons per cubic centimetre (whereas the present matter density
of the universe is only about 3 atoms per cubic metre).

However, despite these and many other successes, the model also has a few
shortcomings. These shouldn't really be seen as failures of the model,
but rather as relevant
questions to which the Big Bang model can provide no answer. I'll briefly
mention two of them. The first arises when instead of analysing all cosmic
microwave background photons together one does the analysis for every
direction of the sky. This was first done by the COBE satellite, and then
confirmed (with increasing precision) by a number of other experiments.
One finds a pattern of very small temperature
fluctuations, of about one part in ten thousand relative to the $2.725$ degrees
Kelvin. It turns out that CMB photons have ceased
interacting with other particles when the universe was about 300000 years old.
After that, they basically propagate freely until we receive them.

Now, temperature fluctuations correspond to density fluctuations:
a region which is hotter than average will also be more dense than average.
What COBE effectively saw was a map (blurred
by experimental and other errors) of the universe at age 300000 years,
showing a series of very small density fluctuations. We belive that these
were subsequently amplified by gravity and eventually led to the structures
we can observe today. The question, however, is where did these fluctuations
come from? At present there are a few theoretical paradigms (each
including a range of particular models) which can explain
this---inflation and topological defects---but they both can claim their
own successes and shortcomings, so the situation is as yet far from clear.
On one hand, the predictions of many inflationary models seem to agree quite
well with observations, but none of these successful models is well-motivated
from a particle physics point of view. On the other hand, topological
defect models are more deeply rooted within particle physics, but their
predictions don't seem to compare so well with observations. One should perhaps
also point out that this comparison may not be entirely fair: inflationary
models are far easier to work with, so the predictions of defect models are
not nearly as well known as those of inflationary models---much work remains
to be done in this area.

The other unanswered question is, surprisingly enough, the contents of the
universe. Obviously, we can only `see' directly matter that emits light, but
it turns out that most of the matter in the universe actually doesn't. For
example, the visible parts of galaxies are thought to be surrounded by much
larger `halos' of dark matter, with a size up to 30 times that of the visible
part. If all this matter were visible, the night sky would look pretty much
like van Gogh's `Starry Night'.

Even though we only have indirect evidence for the existence of this dark
matter, we do have a reasonable idea of what it is. About $5\%$ of the
matter of the universe is visible. Another $5\%$ is invisible `normal
matter', that is baryons (protons and neutrons) and electrons. This is
probably in the form of MACHOS (Massive Compact Halo Objects), such as brown
dwarfs, white dwarfs, planets and possibly black holes. Roughly $25\%$ of
the matter of the universe is thought to be `Cold Dark Matter', that is,
heavy non-relativistic
exotic particles, such as axions or WIMPS (Weakly Interacting Massive
Particles). Cold Dark Matter tends to collapse (or `clump') into the halos of
galaxies, dragging along the dark baryons with it.

Finally, about $65\%$ of the
contents of the universe is thought to be in the form of a `Cosmological
Constant', that is energy of the vacuum---this can also be though of as
cosmic antigravity or the weight of space!
Unlike CDM this never clumps: it tends to make the universe
`blow up' by making it expand faster and faster. In other words, it
forces an accelerated expansion---which, according to recent data, has begun
very recently. This data has been taken
very skeptically by some people. In particular, a period of future acceleration
of the universe, while not posing any problems for cosmologists would be
somewhat problematic for string theory (see \S4) as we know it. However,
this is not a basis for judgement---`data' is not a dirty word,
`assumption' and `conjecture' are dirty words.

These ingredients are needed for cosmological model building. One starts with a
theoretical model, `adds in' cosmological parameters such as the age, matter
contents, and so forth, and computes its observational consequences.
Then one must compare notes with observational cosmologists and see
if the model is in agreement with observation: if it doesn't one had better
start again. In the hope of eliminating some of the shortcomings of the
Big Bang model, one needs to generalise the model, and yet unexplored
extra dimensions are a good place to look for answers.

\section{Strings and extra dimensions}

It is believed that the unification of the known fundamental interactions of
nature requires theories with additional spacetime dimensions. Indeed, the only
known theory of gravity that is consistent with quantum mechanics is String
Theory, which is formulated in 10 dimensions (Polchinski 1998).

Even though
there are at present no robust ideas about how one can go from these theories
to our familiar low-energy spacetime cosmology in four dimensions
(three spatial dimensions plus time),
it is clear that such a process will necessarily involve procedures known
as dimensional reduction and compactification.
These concepts are mathematically very elaborate, but physically quite
simple to understand. Even if the true `theory of everything' is
higher-dimensional, one must find how it would manifest itself to
observers like us who can only probe four dimensions. Note that this is
more general than simply obtaining low energy or other limits of the theory.

On the other hand, given that we only seem to be able to probe four
dimensions, we must figure out why we can't see the others or, in
other words, why (and how) they are hidden. A simple solution is to make these
extra dimensions compact and very small. For example, imagine that you
are an equilibrist walking along a tight rope that is suspended high up in
the air. For you the tight rope will be essentially one-dimensional.
You can safely walk forwards or backwards, but taking a sideways step
will have most unpleasant results. On the other hand, for an ant sitting on
the same rope, it will be two-dimensional: apart from moving forwards and
backwards, it can also safely move around it. It turns out that there are
many different ways of performing such compactifications and, even more
surprisingly, there are ways to make even infinite dimensions not accessible
to us (more on this in \S5).

One of the remarkable general
consequences of these process is that the ordinary
four-dimensional constants become `effective' quantities,
typically being related to the true higher-dimensional fundamental
constants through the characteristic length scales of the
extra dimensions. It also happens that these length scales typically have a
non-trivial evolution. in other words, it is extremely difficult and
unnatural, within the context of string theory,
to make these length scales constant. Indeed, this is such a pressing
question from the string theory point of view
that it has been promoted to the category of a
`problem'---the so called `radius stabilisation problem'. And given
that string theorists are (have to be!) extremely optimistic people,
the fact that they recognise it as being a problem might well be the best
indication that there is something very deep and fundamental about it,
even if at this point we can not quite figure out what it is.

In these circumstances, one is naturally led to the expectation of
time and indeed even space variations of the `effective'
four dimensional constants we can measure.
In what follows we will go through some of the possible cosmological
consequences and observational signatures of these
variations, focusing on the fine-structure constant
($\alpha\equiv e^2/\hbar c$), but also discussing other quantities
in a less extensive way. Before this, however, we need to make a
final excursion into higher-dimensional cosmology.

\section{A cosmological brane scan}

The so-called `brane-world scenarios' are a topic of much recent interest in
which variations of four-dimensional constants emerge
in a clear and natural way.
There is ample evidence that the three forces of particle physics live in
$(3+1)$ dimensions---this has been tested on scales from $10^{-16}$cm to
(for the case of electromagnetism) solar system scales. However, this may
not be the case for gravity. Einstein's field equations have only been
rigorously tested (Will 1993) in the solar system
and the binary pulsar, where the
gravitational field exists essentially in empty space (or vacuum). On smaller
scales, only tests of linear gravity have been carried out, and even so
only down to scales of about two tenths of a millimetre (roughly the
thickness of a human hair).

Sparkled by the existence, in higher-dimensional theories, of membrane-like
objects, the brane world paradigm arose. It postulates that our universe
is a $(3+1)$ membrane that is somehow embedded in a larger space
(commonly called the `bulk') which may or
may not be compact and might even have an infinite volume.
Particle physics in confined (by some mechanism that need not concern
us here) to this brane, while
gravity and other hypothetical non-standard model fields (such as scalar
fields) can propagate everywhere. This may also provide a solution to the
hierarchy problem, that is the problem of
why is gravity so much weaker than any of the other three
forces? The brane world paradigm's answer is simply that this is
because it has to propagate over a much larger volume.

What are possible signatures of extra dimensions? In accelerator physics,
some possible signatures include missing energy (due to the emission of massive
quanta of gravity---gravitons---which escape into the bulk),
interference with standard model processes (new Feynman diagrams with
virtual graviton exchange which introduce corrections to measured
properties such as cross sections), or even more exotic phenomena like strong
gravity effects (such as black holes).

For gravitation and cosmology, the most characteristic sign
would be changes to the gravitational laws, either on very small or very large
scales. Indeed in these models gravity will usually only look four-dimensional
over a limited range of scales, and below or above this range there should be
departures from the four-dimensional behaviour that would be indications for
the extra dimensions. The reason why they appear on small enough
scales can be understood by recalling the tight rope analogy: only something
probing small enough scales (an ant as opposed to the equilibrist) will
see the second dimension. The reason why they should appear on large enough
scale is also easy to understand. If you lived in the south of
England all your life, you could perhaps be forgiven for believing that the
Earth is flat and two-dimensional. However,
once you travel long enough you will start to see mountains, and then
you will realize that it is actually
curved, and hence it must be curved into something---so there
must be some extra dimension for it to curve into.

Other possible clues for brane-type universes and extra dimensions
include changes to the Friedman
equation (for example, with terms induced by the bulk), the appearance of
various large-scale inhomogeneities, and variations
of the fundamental constants---the main topic of this discussion.
Despite this seemingly endless list of possibilities, one should keep in
mind that there are strong observational tests and constraints to be faced,
some of which we already discussed in \S3.

As good example, there are a number of proposals for brane world-type
models where the acceleration of the universe is explained by something other
than a cosmological constant. Such models will have many of
the distinguishing features we just discussed.
A quick search of the literature will reveal about 12 different such models.
Unfortunately, all the models proposed so far fail for
fairly obvious reasons. Since these models will only depart from standard
very recently (usually when acceleration starts) they can fairly easily
be made to agree with the CMB. However, the large-scale changes of gravity
and (in some cases) the additional energy density components, together with
the constraints coming from type Ia supernovae, will make the models run into
trouble when it comes to structure formation: the growth of density
fluctuations, lensing and cluster abundances will all go wrong
(Avelino \& Martins 2002; Uzan \& Bernardeau 2001;
Aguirre \textit{et al.} 2001; White \& Kochanek 2001). Despite this seemingly
disappointing start, brane world scenarios are clearly promising. We simply
don't yet have a clear enough understanding of some of their features,
the most crucial one probably being the interaction mechanisms between
our brane, the bulk , and (if they exist) other branes.

\section{Measuring varying constants: how can we tell?}

So we are now almost ready to start looking for varying constants.
But how would we recognise a varying constant, if we ever saw one?
Two crucial points, which were already implicitly made in \S2 but are worth
re-emphasising here, are that one can only measure dimensionless
combinations of dimensional constants, and that such measurements
are necessarily local.

For example, if I tell you that I am $1.75$ metres tall, what I am really
telling you is that the last time I took the ratio of my height to some
other height which I arbitrarily chose to call `one metre', that ratio
came out to be $1.75$. There is nothing very deep about this number. I would
still be the same height if I had decided to tell you my height in feet and
inches instead. So far so good. Now, if tomorrow I decide to repeat
the above experiment and find a ratio of $1.85$, then that could be either
because I've grown a bit in the meantime, or because I've unknowingly used a
smaller `metre', or due to any combination of these two possibilities.
And the key point is that, even though one of these options might be quite
more plausible than the others, any of them is a perfectly valid description
of what's going on. Moreover, there is no experimental way of
verifying one and disproving the others. Similarly, as regards
the point on locality, the statement that `the speed of light
here is the same  as the speed of light in Andromeda' is either a
definition or it's completely meaningless, since there is no experimental
way of verifying it. These points are crucial and should be
clearly understood (Albrecht \& Magueijo 1999; Avelino \& Martins 1999).

This leads us to an important difference between theory (or model building)
and experiment (or observation). From the theory point of view, it is possible,
and often very convenient, to build models which have varying dimensional
quantities (such as the speed of light, the electron charge or even,
if one is brave enough, Planck's constant). Indeed, such theories became
very popular in recent years.
However, there is nothing fundamental about such choice, in the sense that
any such theory can always be re-cast in a different form,
where another constant will be varying instead, but the observational
consequences of the two will be exactly identical.

For example, given a
theory with a varying constant---say $c$---one can always, by a suitable
re-definition of units of measurement, transform it into another theory
where another constant---say $e$---varies. From our discussion in \S2,
it follows that all we have to do is carry out appropriate re-definitions
of our units of length, time and energy. Again, these two theories
will be observationally indistinguishable,
even though the fundamental equations
may look very different in the two cases---and hence one might strongly
prefer one of the formulations for reasons of simplicity. On the other hand,
the simplest theory that having say a varying $c$, will in general
be different from the simplest theory having a varying $e$, and therefore
these two theories can be experimentally distinguished
(Magueijo \textit{et al.} 2002).

From the observational point of view, it is meaningless to try to measure
variations of dimensional constants \textit{per se}. When considering
observational tests one should focus on dimensionless quantities.
The most relevant example is that of the fine-structure constant,
$\alpha\equiv e^2/(\hbar c)$ which is a measure of the strength of the
electromagnetic interactions.
Other useful parameters are $\beta\equiv G_fm_p^2c/\hbar^3$
and $\mu\equiv m_p/m_e$, where $G_f$ is Fermi's constant and $m_p$ and $m_e$
are respectively the proton and electron masses. Having said this,
we are now ready to begin the search for variations. As we shall see, the
current observational status is rather exciting, but also somewhat confusing.

\section{Local experiments}

Laboratory measurements of the value of the fine-structure constant,
and hence limits on its variation, have been carried out for a
number of years.
The best currently available limit is (Prestage \textit{et al.} 1995)
\begin{equation}
\left|\frac{\dot \alpha}{\alpha}\right|<3.7\times 10^{-14}\3\textrm{yr}^{-1}\, .
\label{labbound}
\end{equation}
This is done by comparing rates between atomic clocks (based on ground state
hyperfine transitions) in alkali atoms with different atomic number Z.
The current best method uses $H$-maser vs $Cs$ or $Hg^+$ clocks,
the effect being a relativistic correction of order $(\alpha Z)^2$.
Future improvements using laser cooled clocks ($Rb$ or $Be^+$)
may improve this bound by about an order of magnitude. Note that this bound
is local, that is, at redshift $z=0$.

On geophysical timescales, the best
constraint (Damour \& Dyson 1996) is
$\left|\dot \alpha/\alpha\right|<0.7\times 10^{-16}\3\textrm{yr}^{-1}$,
although there are suggestions that due to a number of nuclear physics
uncertainties and model dependencies a more realistic bound might be
about an order of magnitude weaker. These come from analysis of $Sm$ isotope
ratios from the natural nuclear reactor at the Oklo (Gabon) uranium mine,
on a timescale of $1.8\times 10^9$ years, corresponding to a cosmological
redshift of $z\sim0.1$.

More recently, Fujii \textit{et al.} (2002) carried out an analysis
of new samples collected at greater depth on the Oklo mine (hoping to minimise
effects of natural contamination). They find two possible ranges of
resonance energy shifts, corresponding to the following values of
$\alpha$
\begin{equation}
\frac{\dot \alpha}{\alpha}=(0.4\pm0.5)\times 10^{-17}\3\textrm{yr}^{-1}\,
\quad \equiv \quad
\frac{\Delta\alpha}{\alpha}=-(0.08\pm0.10)\times10^{-7}
\label{oklobound1}
\end{equation}
\begin{equation}
\frac{\dot \alpha}{\alpha}=-(4.4\pm0.4)\times 10^{-17}\3\textrm{yr}^{-1}\,
\quad \equiv \quad
\frac{\Delta\alpha}{\alpha}=(0.88\pm0.07)\times10^{-7}\, ;
\label{oklobound2}
\end{equation}
note that the first value in each line is an average rate of
change over the period in question; the second is the overall relative change
in the same period. In converting from one to the other one needs to assume
a certain cosmological model---we've assume the standard one, discussed in \S3.
Also note that the latter corresponds to a value of $\alpha$ that was larger
in the past, which might be problematic for some models---see the discussion
in Martins \textit{et al.} (2002).

The authors point out that there is plausible but tentative evidence that
the second result can be excluded by including a further $Gd$ sample.
However, the analysis procedure for $Gd$ data is subject to more uncertainties
than the one for $Sm$, so a more detailed analysis is required before
definite conclusions can be drawn.

It should also be noticed that most theories predicting variations of
fundamental constants can be strongly constrained through gravitational
experiments, most notably via tests of the equivalence principle (Will 1993).

\section{The recent universe}

The standard technique for this
type of measurements, which have been attempted
since the late 1950's, consists of observing the fine splitting of
alkali doublet absorption lines in quasar spectra, and comparing these with
standard laboratory spectra. A different value of $\alpha$ at early times would
mean that electrons would be more loosely (or tightly, depending on the
sign of the variation) bound to the nuclei
compared to the present day, thus changing the
characteristic wavelength of light emitted and absorbed by atoms.
The current best result (Varshalovich \textit{et al.} 2000) using
this method is
\begin{equation}
\frac{\Delta\alpha}{\alpha}=(-4.6\pm4.3\pm1.4)\times 10^{-5}\,,
\qquad z\sim2-4\,;
\label{varshbound1}
\end{equation}
the first error bar corresponds to the statistical (observational)
error while the second is the systematic (laboratory) one. This
corresponds to the bound
$\left|{\dot \alpha}/\alpha\right|<1.4\times 10^{-14}\3\textrm{yr}^{-1}$
over a timescale of about $10^{10}$ years.
They also obtain constraints on spatial variations (by comparing
observations from different regions on the sky),
\begin{equation}
\left|\frac{\Delta\alpha}{\alpha}\right|<3\times 10^{-4}\,, \qquad z\sim2-4\,.
\label{varshbound3}
\end{equation}
Finally, using an analogous technique for $H_2$ lines,
they can also obtain constraints
on the ratio of proton and electron masses,
\begin{equation}
\left|\frac{\Delta\mu}{\mu}\right|<2\times 10^{-4}\,, \qquad z\sim2\,.
\label{varshbound4}
\end{equation}

More recently (Webb \textit{et al.} 2001; Murphy \textit{et al.} 2001)
an improved technique has simultaneously used various multiplets from many
chemical elements to improve the accuracy by about an order of
magnitude. The currently published result is
\begin{equation}
\frac{\Delta\alpha}{\alpha}=(-0.72\pm0.18)\times 10^{-5}\,, \qquad z\sim0.6-3.2\,,
\label{varshbound6}
\end{equation}
corresponding to a four-sigma detection of a {\it smaller} $\alpha$
in the past. Further recent data (Webb 2001, private communication) strengthens
this claim. The analysis of 147 quasar absorption sources (from three
independent data sets obtained with the Keck telescope) provide a six-sigma
detection, $\Delta\alpha/\alpha=(-0.60\pm0.10)\times 10^{-5}$
in the redshift range $z\sim0.6-3.2$. Furthermore,
a large number of tests for systematics have been carried out, and almost all
of these are found not to significantly affect the results. The only two
exceptions are atmospheric disruption and isotopic ratio shifts, but they
act in the `wrong' way: correcting for these would make the detection stronger
(the results presented are uncorrected).

A somewhat different approach consists of using radio and millimetre spectra of
quasar absorption lines. Unfortunately at the moments this can only be
used at lower redshifts, yielding the upper
limit (Carilli \textit{et al.} 2001)
\begin{equation}
\left|\frac{\Delta\alpha}{\alpha}\right|<0.85\times 10^{-5}\,, \qquad z\sim0.25-0.68\,.
\label{varshbound6a}
\end{equation}
Finally, a recent improved technique (Ivanchik \textit{et al.} 2001) of
measuring the wavelengths of $H_2$ transitions in damped Lyman-$\alpha$
systems observed with the ESO VLT/UVES telescope
and using the fact that electron
vibro-rotational lines depend on the reduced mass of the molecule. and this
dependence is different for different transitions, has produced
another claimed possible detection
\begin{equation}
\frac{\Delta\mu}{\mu}=(5.7\pm3.8)\times 10^{-5}\,, \qquad z\sim2.4-3.1\,
\label{varshbound7}
\end{equation}
\begin{equation}
\frac{\Delta\mu}{\mu}=(12.5\pm4.5)\times 10^{-5}\,, \qquad z\sim2.4-3.1\,;
\label{varshbound8}
\end{equation}
here, either value will de bound
depending on which of the two available tables of `standard' laboratory
wavelengths one uses. This clearly indicates that systematic effects aren't
as yet under control in this case.

So a very substantial amount of work has been put into this type of
observations. Even if doubts remain about systematics, a six-sigma detection
is a quite strong result and should be taken seriously. (As a comparison,
the result is roughly as strong as the detection, using Type Ia supernovae
data, of a non-zero value of the cosmological constant.) Now, if these
variations existed at relatively recent times in the history of the
universe, one is naturally led to the question of what was happening
at earlier times---presumably the variations relative to the present day values
would have been stronger then.

\section{The early universe: BBN and CMB}

At much higher redshifts, two of the pillars of standard cosmology
(discussed in \S3) offer very
exciting prospects for studies of variations of constants.
Firstly, BBN has the obvious advantage of probing
the highest redshifts ($z\sim10^{10}$), but
it has a strong drawback in that one is always forced to make
an assumption on how the neutron to proton mass difference depends on $\alpha$.
No known particle physics model provides such a relation, so one usually
has to resort to the phenomenological expression by Gasser \& Leutwyler
(1982), $\Delta m=2.05-0.76(1+\Delta\alpha/\alpha)$.
This is needed to estimate the effect of a varying $\alpha$ on the ${}^4He$
abundance. The abundances of the other light elements depend much less
strongly on this assumption, but on the other hand these abundances are
much less well known observationally.

The cosmic microwave background probes intermediate redshifts,
and has the significant advantage that one has (or will soon have) highly
accurate data.
A varying fine-structure constant changes ionisation history of the
universe: it changes the Thomson scattering cross
section for all interacting species, and also changes the recombination
history of Hydrogen (by far the dominant contribution) and other
species through changes in the energy levels and binding energies.
This will obviously have important effects on the CMB angular power
spectrum, which has now been measured by a number of experiments. Suppose
that $\alpha$ was larger at the epoch of recombination. Then the position of
the first Doppler peak would move smaller angular scales, its amplitude would
increase due to a larger early Integrated
Sachs-Wolfe (ISW) effect, and the damping at small
angular scales would decrease.

Furthermore, a varying $\alpha$ also has an effect on the Large-scale
Structure (LSS) power spectrum, since it changes the horizon size, and
hence the turnover scale in the matter power spectrum. It should be
noticed that although the CMB and LSS are in some sense complementary,
they can not be blindly combined together, since they are sensitive
to different cosmological epochs at which $\alpha$ could have different
values. Therefore the optimal strategy is to use LSS information to
provide priors (in a self-consistent way) for other cosmological
parameters which we can reliably assume to be unchanged throughout the
cosmological epochs in question.

We have recently carried out detailed analyses of the effects of a
varying $\alpha$ on BBN, the CMB and LSS, and compared
the results with the latest available observational results in each
case (Avelino \textit{et al.} 2001, Martins \textit{et al.} 2002).
We find that although the current data has a very slight preference
for a smaller value of $\alpha$ in the past, it is consistent with
no variation and, furthermore, restricts any such variation, from the
epoch of recombination to the present day, to be less than about $4\%$.

The effects of varying constants
are somewhat degenerate with other cosmological parameters, like the baryon
density, Hubble parameter, or re-ionisation. In particular, the effect of the
baryon density seems crucial. For the values quoted above, the baryon density
of the universe agrees with the standard result
(Burles \textit{et al.} 2001). However,
there have been recent claims (Coc \textit{et al.} 2002) that the use
of improved BBN calculations and observations may lead to a lower value of the
baryon density. If one assumes this lower value instead, our estimations
for $\alpha$ would become a detection at more than two sigma.

At a practical level, one needs to find ways of getting around these
degeneracies. Three approaches are being actively tried tried by ourselves
and other groups. The first (obvious) one
is using better data---future CMB satellite experiments such as MAP and
Planck Surveyor should considerably improve the above results, and this
has been recently studied in detail (Martins \textit{et al.} 2002).
The second is using additional microwave background information (such as
polarisation, when data becomes available). And the third and final one is
exploiting the complementarity of various datasets (as hinted above).
While at the moment the constraints on $\alpha$ coming from BBN and CMB data
are at about the 
times be different from the present-day value by about $4\%$,
there is a good chance that progress in both the theoretical understanding
and the quality of the available data will allow us to determine the value of
$\alpha$ from the CMB with much better than about $1\%$ accuracy
within this decade.

\section{So what is your point?}

We have seen that constraints on variations of fundamental constants
at recent times are fairly strong, and any
drastic recent departures from the standard scenario are excluded.
On the other hand, there are no significant constraints in the
pre-nucleosynthesis era, which leaves a rather large open space for theorists
to build models. In between, there are various claimed detections,
particularly from quasar absorption sources at redshifts of a few. These
should definitely be taken seriously, although the situation is far from
settled. The jury is still out on the case of the existence of extra
dimensions: there is as yet no unambiguous observational proof (a
`smoking gun' has not been found),
but considerable supporting evidence is accumulating.

Apart from more observational work, there are deep theoretical issues
to be clarified. The question as to whether all dimensionless parameters in
the final physical `theory of everything'
will be fixed by consistency conditions or if certain of
them will remain arbitrary, is today a question of belief---it does not have
a scientific answer at present. By arbitrary I mean in this context that a
given dimensionless parameter assumed its value in the process of the
cosmological evolution of the universe at an early stage of it. Hence, with s
greater or lesser probability, it could also have assumed other values,
and it could possibly also change in the course of this evolution.

Physics is a logical activity, and hence (unlike other intellectual pursuits),
frowns on radical departures. Physicists much prefer to proceed by
reinterpretation, whereby elegant new ideas provide a sounder basis for what
one already knew, while also leading to further, novel results with at
least a prospect of testability. However, it is often not easy to see how
old concepts fit into a new framework.
How would our views of the world be changed if decisive evidence is
eventually found for extra dimensions and varying fundamental constants?

Theories obeying the Einstein and Strong Equivalence Principles
are metric theories of gravity (Will 1993) In such theories the spacetime
is endowed with a symmetric metric, freely falling bodies follow
geodesics of this metric, and in local freely falling frames the
non-gravitational physics laws are written in the language of special
relativity. If the Einstein Equivalence Principle holds,
the effects of gravity are equivalent to the effects of living
in a curved spacetime. The Strong Equivalence Principle contains the
Einstein Equivalence Principle as the special case where local
gravitational forces (such as Cavendish experiments or
gravimeter measurements, for example)
are ignored. If the Strong Equivalence Principle is strictly valid,
there should be one and only one gravitational field in the universe,
namely the metric.

Varying non-gravitational constants are forbidden by General Relativity and
(in general) all metric theories. A varying fine-structure constant will
violate the equivalence principle, thus signalling the breakdown of
(four-dimensional) gravitation as a geometric phenomenon. It will also
reveal the existence of further (as yet undiscovered) gravitational fields
in the universe, and may be a
very strong proof for the existence of additional spacetime dimensions.
As such, it will be nothing short of a revolution---even more drastic
than the one where Newtonian gravity became part of Einsteinian gravity.
Also, while not telling us, by itself, too much about the `theory of
everything', it will provide some strong clues about what and where to
look for.

Most people (scientists and non-scientists alike) normally make a distinction
between physics (studying down-to-earth things) and astronomy
(studying the heavens above). This distinction is deeply rooted in
pre-historic times, and is still clearly visible today. Indeed, in
my own area of research, such distinction has only started to blur some
thirty years or so
ago, when a few cosmologists noted that the early universe should have been
through a series of phase transitions, of which particle physicists knew a
fair amount about, and hence it would be advisable for cosmologists to
start learning particle physics. Nowadays the circle is closing, with
particle physicists finally beginning to realize that, as they try to
probe earlier and earlier epochs where physical conditions
are more and more extreme, there
is no laboratory on Earth capable of reproducing such these conditions.
Indeed, the only laboratory that is fit for the job is the early universe
itself. Hence it is also advisable for particle physicists to learn cosmology.

The topic of extra dimensions and varying fundamental constants is, to
my mind, the perfect example of a problem at the borderline
between the two areas,
where knowledge of only one of the sides, no matter how deep,
is a severe handicap. This obviously makes it a difficult topic to work
on---but also an extremely exciting one.

\begin{acknowledgements}
The work presented here was done in collaboration with Pedro Avelino,
Rachel Bean, Salvatore Esposito, Gianpiero Mangano, Alessandro Melchiorri,
Gennaro Miele, Ofelia Pisanti, Graca Rocha, Roberto Trotta and
Pedro Viana---I thank them for such an enjoyable and productive
collaboration. I'm also grateful to
Carmen Kachel, Bernard Leong and Carsten van de Bruck for their
comments and suggestions on earlier versions of this article.

This work has been supported by FCT (Portugal)
grant no. FMRH/BPD/1600/2000, and by research projects ESO/PRO/1258/98
and CERN/FIS/43737/2001.
Part of it was performed on COSMOS, the Origin2000 owned by the UK
Computational Cosmology Consortium, supported by Silicon Graphics/Cray
Research, HEFCE and PPARC.
\end{acknowledgements}

\label{lastpage}
\end{document}